# Quantum Gravity :
# Motivations and Alternatives [1]

**Reiner Hedrich**[2]

Institut für Philosophie und Politikwissenschaft
Fakultät Humanwissenschaften und Theologie
Technische Universität Dortmund
Emil-Figge-Strasse 50
44227 Dortmund
Germany

Zentrum für Philosophie und Grundlagen der Wissenschaft
Justus-Liebig-Universität Giessen
Otto-Behaghel-Strasse 10 C II
35394 Giessen
Germany

## Abstract

The mutual conceptual incompatibility between General Relativity and Quantum Mechanics / Quantum Field Theory is generally seen as the most essential motivation for the development of a theory of Quantum Gravity. It leads to the insight that, if gravity is a fundamental interaction and Quantum Mechanics is universally valid, the gravitational field will have to be quantized, not at least because of the inconsistency of semi-classical theories of gravity. The objective of a theory of Quantum Gravity would then be to identify the quantum properties and the quantum dynamics of the gravitational field. If this means to quantize General Relativity, the general-relativistic identification of the gravitational field with the spacetime metric has to be taken into account. The quantization has to be conceptually adequate, which means in particular that the resulting quantum theory has to be background-independent. This can not be achieved by means of quantum field theoretical procedures. More sophisticated strategies, like those of *Loop Quantum Gravity*, have to be applied. One of the basic requirements for such a quantization strategy is that the resulting quantum theory has a classical limit that is (at least approximately, and up to the known phenomenology) identical to General Relativity.

---

[1] Research for this paper was generously supported by the *Fritz-Thyssen-Stiftung für Wissenschaftsförderung* under the project *Raumzeitkonzeptionen in der Quantengravitation*. Thanks also to Brigitte Falkenburg!
[2] Email: Reiner.Hedrich@phil.uni-giessen.de  &  hedrich@fk14.tu-dortmund.de

However, should gravity not be a fundamental, but an induced, residual, emergent interaction, it could very well be an intrinsically classical phenomenon. Should Quantum Mechanics be nonetheless universally valid, we had to assume a quantum substrate from which gravity would result as an emergent classical phenomenon. And there would be no conflict with the arguments against semi-classical theories, because there would be no gravity at all on the substrate level. The gravitational field would not have any quantum properties to be captured by a theory of Quantum Gravity, and a quantization of General Relativity would not lead to any fundamental theory. The objective of a theory of 'Quantum Gravity' would instead be the identification of the quantum substrate from which gravity results. The requirement that the substrate theory has General Relativity as a classical limit – that it reproduces at least the known phenomenology – would remain.

The paper tries to give an overview over the main options for theory construction in the field of Quantum Gravity. Because of the still unclear status of gravity and spacetime, it pleads for the necessity of a plurality of conceptually different approaches to Quantum Gravity.

**Keywords:**
Quantum Gravity, Covariant Quantization, Canonical Quantization, Loop Quantum Gravity, String Theory, Emergent Gravity, Emergent Spacetime, Pregeometry, Quantum Causal Histories

# 1.    Motivations

The most essential motivations for the development of a theory of Quantum Gravity are generally supposed to be based on two (probably interrelated) types of problems: (i) the mutual conceptual incompatibility between General Relativity on the one hand and Quantum Mechanics and Quantum Field Theory on the other hand, and (ii) specific physical problems, unsolved within the framework of the established theories and resulting at least partially from the fact that General Relativity predicts singularities: spacetime points for which it loses its validity.

## 1.1.    The Mutual Conceptual Incompatibility of General Relativity and Quantum Mechanics / Quantum Field Theory

The following three points should elucidate some of the crucial aspects of the conceptual incompatibility between General Relativity and Quantum Mechanics / Quantum Field Theory:[3]

(1) General Relativity, today our best theory of gravity as well as of spacetime, treats the gravitational field as a classical dynamical field, represented by the (pseudo-) Riemannian metric of spacetime.[4] But, according to Quantum Mechanics, dynamical fields have quantum properties. So, if Quantum Mechanics is taken to be universally valid, it seems reasonable to assume the necessity of a (direct or indirect)[5] quantization of the gravitational field. – An additional motivation for the quantization of gravity comes from rather conclusive arguments against semi-classical modifications of the Einstein field equations, i.e. a formalism treating gravity classically and everything else quantum mechanically.[6]

---

[3] Under which conditions this conceptual incompatibility has to be seen as real or as only apparent, as well as what follows from each of these possibilities, will have to be discussed later. See section 2.

[4] All other fields as well as matter are also treated classically by General Relativity.

[5] See sections 2. and 3.1.

[6] Cf. Kiefer (1994, 2004, 2005), Peres / Terno (2001), Terno (2006), Callender / Huggett (2001a, 2001b).

(2) In General Relativity the gravitational field is represented by the metric of spacetime. Therefore, a quantization of the gravitational field would correspond to a quantization of the metric of space-time. The quantum dynamics of the gravitational field would correspond to a dynamical quantum spacetime. But Quantum Field Theories presuppose a fixed, non-dynamical background space for the description of the dynamics of quantum fields. They are conceptually inadequate for a description of a dynamical quantum geometry. An attempt to find a quantum description of dynamical geometry by means of a theoretical approach that necessarily presupposes a background space with an already fixed metric will scarcely be successful. A quantum theory of the gravitational field can scarcely be a Quantum Field Theory, at least not one in the usual sense. – But it is not only the dynamical character of general relativistic spacetime that makes traditional background-dependent quantum theoretical approaches problematic. It is foremost the active diffeomorphism invariance[7] of General Relativity that is fundamentally incompatible with any fixed background spacetime.[8]

(3) In General Relativity, time is a component of dynamical spacetime. It is dynamically involved in the interaction between matter/energy and the spacetime metric. It can be defined only locally and internally; there is no external global time parameter with physical significance.[9] Quantum Mechanics, on the other hand, treats time as a global background parameter, not even as a physical observable represented by a quantum operator.

## 1.2.    Unsolved Physical Problems

Although it is commonly assumed that gravity is a universal interaction and that Quantum Mechanics is universally valid, most physical problems can be captured *either* by General Relativity (e.g. celestial mechanics, GPS positioning) *or* by Quantum Mechanics (e.g. hydrogen atom, electromagnetic radiation). However, there are specific physical situations, in which both of these mutually incompatible conceptual frameworks – General Relativity *and* Quantum Mechanics – would be necessary to get to an adequate description. But such a description can not be achieved because of their mutual incompatibility. Here a theory of Quantum Gravity, by means of which we could get over the mutual incompatibility of General Relativity and Quantum Mechanics, seems to be inevitable. The most prominent of those problematic cases are black holes (*Hawking radiation*[10], *Bekenstein-Hawking entropy*[11]) and the presumed high-density initial state of the universe ('big bang', physics of the early universe, quantum cosmology). In both cases General Relativity predicts singularities; but, because of the breakdown of the equivalence principle for the singularities themselves, the theory becomes inapplicable for these points in spacetime. The fact that General Relativity predicts singularities – points for which it loses its validity – indicates that it can not be a universal theory of spacetime.

According to common wisdom, a successful, adequate theory of spacetime should be able to describe what happens in those cases in which General Relativity predicts singularities. Such a theory – conventionally subsumed under the label 'Quantum Gravity', irrespective of the concrete details – should capture the presumed quantum properties of the gravitational field and of dynamical spacetime. Or it should be able to explain, how gravity and/or spacetime as possibly emergent, intrinsically classical phenomena with no quantum properties could be compatible with – and result from – a quantum world consisting of quantum matter and quantum interactions.[12] It should also explain, which microstates are responsible for the *Bekenstein-Hawking entropy* of black holes; in the classical case, black holes are described by only a few physical quantities that can scarcely be responsible for their (immense) entropy. And a theory of Quantum Gravity should describe the details leading to the *Hawking radiation* of black holes – the details beyond the intuitive quantum field theoretical picture. In particular, it should clarify if *Hawking radiation* leads to a breakdown of the unitarity of Quantum Mechanics – and thereby to an information paradox[13]. And finally it should describe what happens in the final stages of an eventually complete evaporation of a black hole. For all that, it will very probably be inevitable to reach at a description of the black hole event horizon going beyond the classical picture.

---

[7] Cf. Earman (2006, 2006a).
[8] Cf. Earman (1986, 1989, 2002, 2006, 2006a), Earman / Norton (1987), Norton (1988, 1993, 2004).
[9] It is again the active diffeomorphism invariance of General Relativity that leads not at least to the *Problem of Time*. Cf. Belot / Earman (1999), Earman (2002), Pons / Salisbury (2005), Rickles (2005), Rovelli (1991a, 1991b, 2001, 2002, 2006), Isham (1993), Unruh / Wald (1989).
[10] Cf. Hawking (1974, 1975), Bardeen / Carter / Hawking (1973).
[11] Cf. Bekenstein (1973, 1974, 1981, 2000, 2001, 2003), Wald (2001), Bousso (2002). See also section 2.

## 2.      Conceptual Considerations

The well-established, empirically well-confirmed precursor theories – General Relativity and Quantum Mechanics –, together with the already existing empirical data that confirmed these theories, are still the only concrete elements that constitute a reasonable starting point for the different attempts to construct a theory of Quantum Gravity, intended to get over their mutual conceptual incompatibility. There are still no relevant empirical data that point without doubt beyond those precursors. In this situation, the most fundamental requirements for theory construction in Quantum Gravity are, on the one hand, conceptual coherence and consistency. On the other hand it is the necessity to reproduce at least the empirical basis of the well-established theories – their phenomenology –, which means that theoretical approaches in Quantum Gravity have to reproduce those precursors at least as approximations or low-energy implications.[14]

The freedom left for theory development, after taking into account (or at least having the intention to take into account) those basic requirements, is usually filled by (sometimes rather problematic) metaphysical assumptions. Which basic conceptual (or model-theoretical) elements of the established precursor theories – beyond their phenomenology – are taken to be essential for the development of the new theoretical approaches depends primarily on the assessment of those elements with regard to their relevance for Quantum Gravity. Because of the conceptual incompatibility of the precursor theories it has necessarily to be a selection. And there are no objective a priori criteria for this selection. Idiosyncratic convictions enter at this point. – Is the background-independence of General Relativity indeed to be seen as a basic conceptual requirement for Quantum Gravity? Is spacetime fundamental or emergent? Is it a substance or a relational construct? If it is a substance, does it have quantum properties? Is spacetime based on (or does it emerge from) a quantum substrate or rather something completely different? Has the theory of 'Quantum Gravity' necessarily to be a quantum theory? Has the fundamental theory to be a nomologically or ontologically unified theory?

So, with this caveat in mind, what could be reasonable elements of a starting point for the development of a theory of Quantum Gravity? What should be taken as at least heuristically relevant? Which conceptual elements of the precursor theories constitute presumably essential physical insights that will probably survive the next step in the development of a coherent and empirically adequate picture of physical reality? What should at least be taken into account?

---

[12] See sections 3.2. and 3.4.

[13] Cf. Hawking (1976, 1982, 2005), Belot / Earman / Ruetsche (1999).

[14] Even better would be predictions that contradict in very specific details the established theories, but that do not contradict already existing empirical data. Such predictions could lead to a perspective for differential empirical tests.

One of the most fundamental insights of General Relativity – our empirically well-confirmed classical theory of gravitation and of spacetime – is that it is the metric of spacetime which represents the gravitational field. If we take this geometrization of gravity seriously, that means that the gravitational field is (unlike all other interaction fields) not a field defined on spacetime, but rather a manifestation of spacetime itself. Consequently, it is not possible to describe the dynamics of the gravitational field on an already predefined (or fixed) background spacetime. As long as there are no better, well-founded reasons, a theory of Quantum Gravity has to take into account this background-independence; it has to describe the dynamics of the gravitational field without recourse to an already existing spacetime (metric). Additionally, under extrapolation of the conceptual implications of General Relativity, one could suspect, at least for the time being, that a successful theory of Quantum Gravity will probably not only be a theory describing a *dynamical* spacetime, rather it will be based on a *relational* conception of spacetime[15] – or it will even lead to an emergent spacetime scenario.

If we take Quantum Mechanics seriously as our fundamental (and presumably universally valid) theory of the dynamics of matter and fields, it seems to be reasonable (at least at first sight) to assume that the gravitational field – like all other dynamical fields – should have quantum properties. Much more clearly than this intuition, the arguments against semi-classical theories of gravitation exclude the possibility of a fundamental non-quantum gravitational interaction in a quantum world.

But this does not exclude the possibility that gravity – in contrast to the intuition leading to the assumption of quantum properties of the gravitational field – could be an intrinsically classical phenomenon, emerging from a quantum substrate without gravitational degrees of freedom. It is at least not completely unreasonable to take this possibility into account. Then, gravitation would not be a fundamental interaction; it would be a residual interaction, caused by non-gravitational interactions and their corresponding degrees of freedom. This would not lead to any conflict with the arguments against semi-classical theories, because, on the fundamental level, there would be only the quantum substrate, governed by fundamental quantum interactions, to which gravity would not belong. A theory describing the dynamics of the gravitational field would be an effective theory describing the intrinsically classical dynamics of collective degrees of freedom that result from a completely different quantum substrate; this classical theory would have to be recovered from the fundamental theory by means of a statistical approximation over the (more) fundamental degrees of freedom of the substrate.

However, should gravity indeed be a fundamental interaction (and should Quantum Mechanics be universally valid), then we had to expect for the gravitational field, as a fundamental entity, quantum properties, not yet taken into account in the classical picture provided by General Relativity. The gravitational field would have to be 'quantized' – like the electromagnetic field, but very probably not with the model-theoretical apparatus used in Quantum Electrodynamics, because of the fixed background spacetime necessarily required in Quantum Field Theory.

---

[15] Because of the active diffeomorphism invariance of General Relativity, that can be understood as a gauge invariance (cf. Earman (1986, 1989, 2002, 2006, 2006a), Earman / Norton (1987), Norton (1988, 1993, 2004); active diffeomorphisms are to be understood as point transformations, in contrast to passive diffeomorphisms: mere coordinate transformations; so, General Relativity is invariant under transformations of the points of the spacetime manifold), it seems to be highly unreasonable to interpret the spacetime manifold as a substantial entity; the prize for that would consist in rather unmotivated metaphysical assumptions: (i) the negation of *Leibniz equivalence* (i.e. the negation of the identity of the indistinguishable: empirically completely indistinguishable models of spacetime would have to be seen as representations of different spacetimes), and (ii) a completely unmotivated (and unobservable) indeterminism of the theory (as a consequence of the hole argument; cf. the references above). What remains without a substantially interpretable spacetime manifold is: a metric field (identical with the gravitational field; carrying energy and momentum like all other fields), the other interaction fields, the matter fields, and the relations between these fields.

Should General Relativity be the adequate classical theory to be quantized and should it capture the relevant classical features of gravity, then its identification of gravity with properties of a dynamical geometry would probably mean that a quantization of the gravitational field corresponds to a quantization of dynamical spacetime. The quantization of gravity would lead to a quantum geometry. At least on first sight, one could suspect that a (conceptually and empirically successful) quantization of General Relativity – should it be achievable – would lead to a theory describing the metric of spacetime as an expectation value of a quantum variable; furthermore one would probably expect something like uncertainties of spacetime, or quantum fluctuations of the spacetime metric, of spacetime geometry, possibly even of the spacetime topology.

But this strategy for the development of a theory of Quantum Gravity, i.e. constructing it by means of a (direct) quantization of General Relativity, intended to identify and capture the quantum properties of gravity and spacetime, will only be successful if gravity is indeed a fundamental interaction, if the gravitational field (as well as spacetime) has indeed quantum properties. If gravity is an intrinsically classical phenomenon, this strategy will simply lead to a quantization of the wrong degrees of freedom: macroscopic collective degrees of freedom resulting from a totally different substrate, governed by totally different fundamental degrees of freedom, which then should be the original subject of a theory of 'Quantum Gravity' – supposed that one has the intention to keep up this name for the theory by means of which we would get over the conceptual incompatibility between General Relativity and Quantum Mechanics, irrespective of the question if it is a factual or only an apparent incompatibility.

So, under consideration of the possibility that gravity could either be a fundamental interaction or an intrinsically classical phenomenon – which means to take all possibilities into account – 'Quantum Gravity' would (irrespective of the details) be the name of the theory by means of which we are able to explain the dynamics (and possibly the emergence) of gravity (and spacetime) in a way that gets over the (factual or only apparent) conceptual incompatibility between General Relativity and Quantum Mechanics. It would be a theory describing the substrate of gravity (and spacetime). And this substrate may either contain (quantum) gravitational degrees of freedom or not. The options are still open: fundamental or emergent gravitational interaction, fundamental or emergent spacetime, quantum geometry or intrinsically classical spacetime, substantial or relational spacetime, etc. Taking all options seriously, the theory of 'Quantum Gravity' we are searching for should be a quantum theory (in the broadest sense[16]), which – and this is the most basic requirement for any such theory – reproduces the phenomenal content of General Relativity (at least approximately and without conflict with known empirical data; possibly as a classical, macroscopic, low-energy limit), and which should be able to explain the empirical and conceptual successes of General Relativity. Additionally, at least on the long run, such a theory has to lead to specific own prediction that go beyond those of its precursor theories, leading thereby to specific and differential forms of experimental testability.

This minimum definition (and broad conception) of 'Quantum Gravity' leads to a wide spectrum of options for theory development. The already existing approaches[17] differ especially with regard to the specific conceptual and model-theoretical components of the precursor theories they take to be essential for Quantum Gravity or as indispensable for its modalities of theory construction. But, here again, it has to be emphasized that the probably inevitable inclusion of conceptual elements derived from the precursor theories is not completely unproblematic. The well-established precursors and their conceptual implications could just point towards the wrong direction. Taking these precursor theories more or less uncritically as constitutive components of a starting point for an attempt to eradicate their mutual conceptual incompatibility – the dominant attitude at least for the direct quantization approaches[18] – could possibly lead into dead ends. A careless extrapolation of elements from the precursor theories might be fatal and should at least not be taken as the only option for the theory development in Quantum Gravity. On the other hand, too speculative approaches, far from the conceptual basis of the precursor theories, bear without doubt their own risks. Only a pluralistic strategy in theory construction will be adequate to make these different risks of every single approach more controllable. All reasonable alternatives should be taken into account, even those alternatives that, on first sight, seem to be eccentric in comparison with the standards of our well-established theories. Such eccentric alternatives could nonetheless be able to reproduce the implications of these precursors as low-energy approximations and lead to specific new predictions – and, thereby, to empirical testability.

---

[16] Although this is finally nothing more than a question of nomenclature, fundamental non-quantum theories are here formally excluded from the spectrum of theories that go by the name 'Quantum Gravity'. For such approaches to a fundamental theory, see section 4.1.

[17] See section 3.

There is still something that was not yet mentioned with regard to the prerequisites of a possible starting point for the theory development in the field of Quantum Gravity, something that could nonetheless be of specific heuristic significance in this context. The statement made at the beginning of this section, that General Relativity and Quantum Mechanics are still delivering the only concrete conceptual elements for a starting point with regard to the attempts to get over their mutual conceptual incompatibility, is not quite correct. Additionally, there are conceptions and ideas that are motivated sufficiently well within the context of our established theories, but do not belong any more fully to this context – something that could be called 'elements of transition'. What makes these 'elements of transition' interesting and relevant for Quantum Gravity, is that they already constitute bridges going over the frontiers between the otherwise conceptually incompatible established theories.

A paradigmatic example of such an 'element of transition', conceptually going beyond the context of our established theories, concerns the *Bekenstein-Hawking entropy*[19] of black holes. It results from considerations that combine implications of General Relativity, Quantum Mechanics, Thermodynamics and Information Theory. And it is this combination that makes black hole entropy interesting and relevant for Quantum Gravity. The *Bekenstein-Hawking entropy* of black holes, and the *Covariant* (or *Holographic*) *Entropy Bound*[20] which can be motivated within the thermodynamics of black holes, point directly to a discrete substrate at the Planck level. This can be taken as an indication *either* of a discrete spacetime structure – if spacetime should be fundamental – *or* of a discrete structure from which spacetime emerges. In any case, it is pointing directly to a substrate with a finite number of degrees of freedom per spacetime region – irrespective of the question, if this spacetime region is part of the fundamental level or emergent, i.e. if the fundamental degrees of freedom are (quantum) geometric or 'pregeometric'. Additional, but much more indirect indications of such a discrete structure come from the singularities that General Relativity predicts (but which transcend its model-theoretical apparatus: differential geometry) and from the divergences that occur in Quantum Field Theory for small distances / high energies. Both could be artifacts of the continuum assumption with regard to spacetime or of the assumption of an infinity of the relevant degrees of freedom respectively.

Interestingly, almost all existing approaches to a theory of Quantum Gravity lead to indications either of a discrete spacetime structure – for those approaches that take spacetime to be a funda-mental entity whose quantum properties have to be revealed in the context of a theory that goes beyond General Relativity in exactly this point – or of a discrete ('pregeometric') substrate structure from which spacetime results. What is of specific significance, is that these indications of a discrete substructure are not only present in the more radical approaches, but also in those approaches that take the fundamentals of General Relativity as well as those of Quantum Mechanics to be essential for a theory of Quantum Gravity. This is most astonishing, because the assumption of a spacetime continuum and of an infinite number of physically relevant degrees of freedom is an inevitable ingredient of General Relativity (differential geometry presupposes the continuum) as well as of Quantum Mechanics and Quantum Field Theory (fields are defined on a classical continuous background space). The best example of a very sophisticated, but at the same time very conservative approach to Quantum Gravity that takes the conceptual basis of General Relativity as well as that of Quantum Mechanics very seriously as essential part of its conceptual strategy, but nonetheless leads directly to a discrete structure when General Relativity is quantized, is *Loop Quantum Gravity*.[21]

---

[18] See section 3.1.
[19] Cf. Bekenstein (1973, 1974, 1981, 2000, 2001), Wald (1994, 2001), Bousso (2002). See also section 1.2.
[20] Cf. Bekenstein (1981, 2000, 2001), Bousso (2002), Pesci (2007, 2008).

## 3. Strategies for the Development of a Theory of Quantum Gravity

### 3.1. Quantization of General Relativity

Considering the direct quantization of General Relativity as a reasonable strategy to overcome its (apparent) conceptual incompatibility with Quantum Mechanics and Quantum Field Theory, one has to remember that the most essential requirement for a theory of Quantum Gravity consists – besides conceptual consistency and coherence – in its ability to reproduce General Relativity after its quantization as a classical limit or low-energy approximation of the quantized theory. The quantum theory has at least to reproduce the macroscopic phenomenology of its classical starting point up to the exactitude of the already existing empirical data. We will see in the following that this requirement is not necessarily or automatically fulfilled. A quantization of General Relativity does not necessarily lead back to it. But the way back is the essential requirement for a theory of Quantum Gravity – and not that it was constructed by a quantization of the empirically well-confirmed classical theory.

If one tries nonetheless to follow the direct quantization strategy, one has to decide which quantization procedure should be applied; there are different methodological options, even when the classical theory is well-known and well-defined. Furthermore one has to decide which physical magnitude has to be quantized. In the case of General Relativity, it could be the metric, or topology, or even the causal structure. And for all these decisions the question remains how to take into account the background-independence of the classical theory during quantization. – As we will see, all existing direct quantization approaches start from a quantization of the metric or of physical magnitudes on the same descriptive level: connections, holonomies. Their most striking difference is to be found in their respective attitude with regard to background-independence and its formal realization.

### 3.1.1. *Covariant Quantization*

The *Covariant Quantization*[22] of General Relativity consists in the attempt to construct a Quantum Field Theory of gravity, which means: a Quantum Field Theory of the metric field, in the manner of Quantum Electrodynamics, the Quantum Field Theory of the electromagnetic field. But Quantum

---

[21] See section 3.1.2.
[22] Cf. DeWitt (1967a, 1967b)

Field Theories in this orthodox sense need a background spacetime with a fixed metric for the definition of its operator fields. Consequently, *Covariant Quantization* uses a standard perturbation-theoretical approach, working with a fixed kinematical (usually Minkowski) background metric and a perturbation on this background to be treated quantum mechanically. This leads to a Quantum Field Theory of the fluctuations of the metric. The corresponding field quanta of gravity, called 'gravitons', are massless and have spin 2 – as a consequence of symmetry arguments and of the properties of classical gravity (long-range, exclusively attractive). They are assumed to represent the quantum properties of spacetime and to behave according to standard Feynman rules on a fixed background spacetime.

But *Covariant Quantization* with its perturbation expansion of the fluctuations of the spacetime metric turns out to be non-renormalizable. This is a direct consequence of the self-interaction of the graviton, which, in turn, is nothing else than a quantum-field-theoretical expression of the nonlinearity of classical gravity. Gravity couples to mass and, because of the mass-energy equivalence, to every form of energy.[23] Therefore the self-interaction contributions to gravity increase for decreasing distances or increasing energies. So, the contribution of virtual particles with increasing energies dominates the higher orders of the perturbation expansion. This leads to uncontrollable divergences of the expansion and to its non-renormalizability. No quantitative predictions can be achieved. This makes the theory irrelevant as a fundamental description of spacetime.

Anyway, the non-renormalizability of *Covariant Quantization* is not much of a surprise. The background-independence of General Relativity, which is a consequence of the identification of gravity with properties of spacetime itself, makes a background-dependent approach to a theory of Quantum Gravity highly questionable. *Covariant Quantization* tries to quantize a background-independent theory – General Relativity – by means of a necessarily background-dependent method: obviously a conceptual contradiction. At the same time, it shows that it is not possible to get over the mutual conceptual incompatibility between General Relativity and Quantum Mechanics / Quantum Field Theory by simply amalgamating gravity and the quantum by means of the standard quantization procedures. The conceptual foundations of both are obviously much too different. A Quantum Field Theory (in the orthodox sense) of gravity does not exist, because it is not possible to quantize a background-independent theory of spacetime by means of a background-dependent approach, describing the quantum dynamics of spacetime on (an already fixed) spacetime.

### 3.1.2. *Canonical Quantization – Loop Quantum Gravity*

The *Canonical Quantization* of General Relativity does not lead to *these* problems. It is a much more sophisticated, intrinsically non-perturbative, background-independent, full-blown quantization of General Relativity, starting from its Hamiltonian formulation. Nonetheless the old geometrodynamical[24] *Canonical Quantization* approach, which started from a quantization of a Hamiltonian formulation of General Relativity with the metric and the curvature of spacetime as basic variables, led to severe and probably insoluble problems. Its fundamental equation, the *Wheeler-Dewitt equation* (i.e. the quantized counterpart to the *Hamiltonian constraint* of the classical theory that captures the temporal aspect of its diffeomorphism invariance[25]), turned out to be ill-defined and led to severe conceptual and mathematical problems. Therefore we will turn over directly to *Loop Quantum Gravity*, the more successful reincarnation of the *Canonical Quantization* approach.

---

[23] All other interactions couple only to their 'charges', not to energy.
[24] Cf. DeWitt (1967), Kuchar (1986, 1993), Ehlers / Friedrich (1994).
[25] See below.

*Loop Quantum Gravity*[26] starts from a Hamiltonian formulation of General Relativity based on the *Ashtekar variables*[27] (a spatial SU(2) connection variable and an orthonormal triad) instead of the metric and the curvature of spacetime as basic variables. – The Hamiltonian formulation of General Relativity results from a splitting of spacetime into spatial hypersurfaces and a time parameter. In the case of the Ashtekar variables, it is a three-dimensional connection and a time parameter. The latter is necessary for the definition of the canonical momentum as well as for the canonical quantization procedure. The (active[28]) diffeomorphism invariance of General Relativity – the formal expression of the general covariance of the classical theory, interpreted in *Loop Quantum Gravity* as a gauge invariance[29] (that has to be taken into account in the transition to the quantum theory) – translates in the Hamiltonian approach to the *constraints*.[30] These constraints are necessary, because the plain Hamiltonian theory and its basic variables do not take into account diffeomorphism invariance. The corresponding full phase space contains redundant representations of physically identical spacetimes (as well as representations of physically impossible states – states that lie outside the 'constraint surface'). The identification of equivalence classes of representations of physically identical spacetimes – equivalence classes of representations that can be transformed into each other by a diffeomorphism – (as well as the identification of physically impossible states) has to be introduced additionally, by means of the constraints.

Constraints are typical for the Hamiltonian formulation of dynamics with an unphysical surplus structure. Such an unphysical surplus structure is, on the other hand, typical for systems with gauge freedom. In gauge systems, it is the gauge invariance that captures unphysical redundancies in the description of a system, in the Hamiltonian formalism it is the constraints that capture them. The constraints can be understood as generators of gauge transformations. In General Relativity the corresponding gauge invariance is diffeomorphism invariance. – Gauge transformations are unobservable, and if one wants to keep up the predictive power of the theory, then 'observables' have to be gauge-invariant. Formally, in the Hamiltonian approach, this means that all observables have (weakly, i.e. on the constraint surface) vanishing Poisson brackets with all (first class[31]) constraints. In the quantum case this translates into: all quantum observables have to commute with all quantum constraints.

Already in the geometrodynamical version of the Hamiltonian formulation of General Relativity, after the splitting of spacetime into spatial hypersurfaces and a time parameter, there are four constraints: the *scalar* or *Hamiltonian constraint* and three *momentum* or *diffeomorphism constraints*.[32] In the Ashtekar version, because of an additional redundancy connected with the new variables, one has three additional *Gauss constraints* that generate SU(2) gauge transformations. *Loop Quantum Gravity* starts from here, after a further technical modification of the classical Hamiltonian theory – a transition from Ashtekar's connection variables to loop variables (Wilson loops)[33] – into the quantization procedure, using the Dirac quantization method[34] for constrained Hamiltonian systems.

---

[26] Cf. Ashtekar (2007, 2007a), Ashtekar / Lewandowski (2004), Ashtekar et al. (1992), Rovelli (1991, 1991c, 1997, 1998, 2003, 2004), Smolin (1991, 2000), Thiemann (2001, 2002, 2006), Nicolai / Peeters (2006), Nicolai / Peeters / Zamaklar (2005). For a literature survey see Hauser / Corichi (2005).

[27] Cf. Ashtekar (1986, 1987).

[28] See footnote 15.

[29] Spacetimes that can be transformed into each other by means of a diffeomorphism are indistinguishable; they are *Leibniz equivalent*. This is the basis for interpreting active diffeomorphism invariance as a gauge invariance. Only gauge-invariant quantities are understood as constituting the real physical content of the theory; everything else is a model theoretical artifact. Gauge theories contain unphysical redundancies or surplus structures that make the description more indirect, but at same time more practical or more symmetrical. Sometimes there is not even a real alternative to such a redundant description; for General Relativity no formulation based only on the real physical degrees of freedom is known. See below.

[30] Cf. Henneaux / Teitelboim (1992), Govaerts (2002), Belot / Earman (1999, 2001). The (primary) constraints are a direct consequence of the transition from the Lagrangian formalism to the Hamiltonian formalism by means of a Legendre transformation.

[31] First class constraints are constraints with vanishing Poisson brackets with all other constraints.

[32] The momentum constraints capture the invariance under spacelike diffeomorphisms, the Hamiltonian constraint that under timelike diffeomorphisms.

The Dirac quantization method consists in a quantization of the full, unconstrained Hamiltonian phase space of the classical theory – canonical commutation relations for the quantum counterparts of the classical variables, an operator algebra, and finally, the quantum counterparts of the classical constraints are to be defined – with the intention to take the quantum constraints into account (to 'solve the constraints') afterwards, and to identify thereby the true physical states. An alternative to Dirac quantization would consist in solving the constraints first, for the classical theory, and then to quantize the reduced classical theory, which, then, has no constraints any more. Under 'solving the constraints' one understands in the classical case a transition from a description based on the full (unconstrained) Hamiltonian phase space, containing descriptive redundancies (as well as physically impossible states), to a reduced phase space that captures only the 'true' (physical) degrees of freedom of the system. In the quantum case this corresponds to the transition from the full (unconstrained) 'kinematical' quantum mechanical Hilbert space, containing redundancies (in form of gauge symmetries), to a reduced 'physical' Hilbert space representing only the 'true' physical states of the system. – But, actually, the alternative to Dirac quantization is, unfortunately, nothing more than a chimera, because no one knows how to construct the reduced physical phase space of General Relativity; it is generally taken to be impossible. Already at this point, one could ask: Why should it be easier to solve the constraints in the quantum case? And indeed, solving *all* the quantum constraints and finding the physical Hilbert space, and thereby the true states of *Loop Quantum Gravity*, is anything but easy. Actually, no one knows how to do that either. The quantized form of the Hamiltonian constraint, the *Wheeler-DeWitt equation*, is well-known for its resistance against any attempt to solve it. Here *Loop Quantum Gravity* does not offer better solutions than the old geometrodynamical approach.

Nonetheless, there are already very interesting results for the kinematical Hilbert space in *Loop Quantum Gravity*. For the spatial hypersurfaces, after solving only the quantum Gauss constraints, one finds a discrete, polymer-like graph structure: according to *Loop Quantum Gravity*, the discrete quantum substructure to (the spatial part of) the spacetime continuum of General Relativity.[35] This *spin network* structure represents the discrete eigenvalues of two geometric operators one can define in *Loop Quantum Gravity*: the *area* and the *volume operator*. Up to this point, only the Gauss constraints are solved. The spin networks, as well as the related area and volume operators, are not diffeomorphism-invariant; they do not commute with the other quantum constraints. The next step consists in solving the (spatial) diffeomorphism (or momentum) constraints. This is realized in a transition from the spin networks to the diffeomorphism-invariant *S-knots*: equivalence classes of spin networks with regard to spatial diffeomorphisms. S-knots are abstract topological objects – according to *Loop Quantum Gravity*: excitation states of the gravitational field – that do not live on a background space, but rather represent space itself. Although the spacetime manifold is required to derive the S-knots, they are, according to *Loop Quantum Gravity*, the entities defining space. Every localization is a localization with regard to the S-knots. – But S-knots represent only quantum space, not spacetime. They are not invariant with regard to temporal diffeomorphisms. They are not yet the states of the true, physical Hilbert space of the theory. The necessary last step would consist in solving the quantum Hamiltonian constraint (i.e. the Wheeler-DeWitt equation). But, as yet, *Loop Quantum Gravity* has not succeeded with this project.[36] Not even the definition of the quantum Hamiltonian constraint is unambiguous.

---

[33] A Wilson loop is the trace of a holonomy (an integral of a connection along a closed curve). Wilson loops have the advantage of being gauge-invariant; the corresponding holonomies are not necessarily gauge-invariant.

[34] Cf. Henneaux / Teitelboim (1992).

[35] It has to be emphasized that the discreteness of the spin network of *Loop Quantum Gravity* is a result of the direct non-perturbative quantization of General Relativity, not a feature the theory started with. However, the discreteness of the spin networks is not that of a regular cellular arrangement or grid (like e.g. in cellular automata), but a discreteness that requires the continuum of real numbers (like Quantum Mechanics) for its definition. It presupposes the spacetime manifold of General Relativity, although *Loop Quantum Gravity* tries to discuss away the manifold after quantization.

And there are further serious problems in *Loop Quantum Gravity*: One of these, and probably the most severe, is that no low-energy approximation and no classical limit have been derived as yet. It has not been possible to reproduce the known low-energy phenomenology of gravity or to derive the Einstein field equations (or anything similar to them) as a classical limit. – And it is exactly at this point, that one should remember that it is not a necessary requirement for a theory of Quantum Gravity to quantize General Relativity in a conceptually coherent way (although this might be a natural strategy). Rather, the basic and indispensable requirement for such a theory is that it is able to reproduce the phenomenology of gravity: the classical, low-energy case. Should it not be possible to do this, this would be the end of *Loop Quantum Gravity*.[37] So it seems at least to be reasonable to look for alternatives.

### 3.1.3. *Quantization of a Discretized Version of General Relativity*

Besides the *Covariant* and the *Canonical Quantization* schemes, there are at least three approaches that try to quantize General Relativity more or less directly, but in a already discretized form: the *Consistent Discretization*[38], the *Regge Calculus*[39], and the *Dynamical Triangulation* approach[40], the latter in its Euclidean and Lorentzian versions. Those approaches differ with regard to their method of discretization of the classical theory, with regard to the signature of spacetime already presupposed, and especially with regard to their respective success in the attempt to quantize the discretized classical dynamical structure and to reproduce a classical limit compatible with known phenomenology.

---

[36] Some insiders do not even expect (any more) a complete solution to this problem: Only treatments with many simplifications exist.

[37] Should *Loop Quantum Gravity* instead finally be able to master these problems and succeed in the reproduction of the phenomenology of gravity, it is already clear that it has radical implications in comparison to the well-established theories of physics. Its dynamics does not fulfill unitarity and all observables are non-local. The probably most radical of its consequences is the *problem of time* (cf. Belot / Earman (1999, 2001), Butterfield / Isham (1999), Kuchar (1991, 1992), Rickles (2004, 2005), Rovelli (1991a, 1991b, 1998, 2002, 2007), Unruh / Wald (1989)). The *problem of time* is already present in General Relativity, but it has more severe implications in *Loop Quantum Gravity*:
In General Relativity, coordinate time is not diffeomorphism-invariant. According to the gauge-theoretical interpretation of (the Hamiltonian formulation of) General Relativity (cf. Earman (2006, 2006a), Belot / Earman (1999, 2001)), it is a gauge variable. The Hamiltonian constraint, capturing the transition from one spatial hypersurface to another, and therefore the dynamics of the system, can be understood as a gauge transformation. Essentially, this is nothing more than a circumscription of the fact that, as a result of the diffeomorphism invariance of the theory, dynamical transitions, generated by the Hamiltonian constraint, do not lead to any observable consequences. – So, because it is not diffeomorphism-invariant, coordinate time is unobservable in General Relativity. And clock time, as an observable physical quantity, is a non-trivial function of the gravitational field, leading to such effects as the clock paradox. There does not exist any external, observable time parameter in General Relativity. This is finally a consequence of general covariance, captured in the diffeomorphism invariance of the theory.
In the classical case, if we consider only individual solutions of the Einstein equations, the practical consequences of the *problem of time* are limited; in most cases a proper time can be defined for worldlines. This is different for the quantum case because of superpositions of spacetimes and the nonexistence of trajectories. After the canonical quantization of General Relativity there are no fundamental equations that describe a temporal evolution of the system. This is again because of the fact that the temporal evolution of the system is coded into the Hamiltonian constraint, which generates a gauge transformation. The corresponding gauge symmetry reflects nothing more than a descriptive redundancy of the theory, something with no observable physical counterpart. So, the quantized Hamiltonian constraint makes *Loop Quantum Gravity* a theory without time. All observables of the theory are timeless, because all corresponding quantum operators have to commute with the quantum Hamiltonian constraint, into which the temporal evolution of the system is coded. And, apparently, nothing can change this fact, if one is decided only to accept observable quantities as physically relevant, in other words, if one is decided only to accept gauge-invariant operators: quantum observables that commute with all quantum constraints. For a possible solution of this problem see Rovelli (2002).

[38] Cf. Gambini / Pullin (2003, 2004, 2005, 2005a, 2005b), Gambini / Porto / Pullin (2003).

Methodologically, these approaches can be compared more or less with lattice gauge theories. Without any further physical motivation, they can scarcely be taken as an adequate description of the true quantum substrate of gravity and spacetime. Even Lorentzian *Dynamical Triangulation*[41] – the only approach in this spectrum that is not only successful with regard to the quantization of the discretized classical theory, but leads also to results compatible with known phenomenology – should, finally, be seen as nothing more than an effective theory, leading to an adequate low-energy picture. It should not be mistaken for a fundamental theory describing the quantum substrate of spacetime.

### 3.1.4. *Theory Extension before or after the Quantization of General Relativity*

An additional option that could be considered as a possible road to an adequate theory of Quantum Gravity consists in a weakening of the directness of a quantization of General Relativity by means of an extension: One can either extend General Relativity before its quantization or extend the quantum theory resulting from a quantization of General Relativity. Usually, the symmetries of the theory are taken to be the main object of such an extension. To this context belongs a theory called *Supergravity*[42] – a supersymmetric and, because of consistency requirements, eleven-dimensional extension of a quantum version of General Relativity. In the seventies and eighties, *Supergravity* was taken very seriously as a promising option for a theory of Quantum Gravity. This perspective vanished with the discovery of conceptual problems and increasing doubt with regard to the renormalizability of the theory. Finally, it came to a resurrection of the approach as an effective theory; *Supergravity* aroused the interest of string theoreticians, who found a new role for it as part of the web of dualities between the perturbative string theories.[43]

### 3.2. Conditions leading to the Inadequacy of a Quantization of General Relativity – Specification and Exemplification

As already mentioned in section 2, there is the possibility that gravity could be an intrinsically classical, macroscopic phenomenon. As also mentioned before, an intrinsically classical gravity does not lead to conflicts with the arguments against semi-classical theories of gravity, if it is an emergent phenomenon, resulting from a quantum substrate that does not contain any gravitational degrees of freedom. The arguments against semi-classical theories of gravity presuppose that gravity is a fundamental interaction. They lose their validity if gravity is not fundamental, if gravity does not even appear in a fundamental quantum description of nature. Then, on the fundamental level, there would be no semi-classical hybrid dynamics that leads to conceptual inconsistencies. So, if gravity is an intrinsically classical phenomenon, it can not be a fundamental interaction. It has to be an induced or residual effect, caused by a quantum substrate dominated by other interactions.[44]

---

[39] Cf. Regge / Williams (2000), Williams / Tuckey (1992), Gentle (2002), Barrett (1987).

[40] Cf. Ambjorn (1995) Ambjorn / Jurkiewicz / Loll (2000, 2001, 2001a, 2004, 2005, 2005a, 2005b, 2006), Ambjorn / Loll (1998), Loll (1998, 2001, 2003, 2007), Loll / Ambjorn / Jurkiewicz (2005).

[41] Cf. Ambjorn / Jurkiewicz / Loll (2000, 2001, 2001a, 2004, 2005, 2005a, 2005b, 2006), Ambjorn / Loll (1998), Loll (2001, 2003, 2007), Loll / Ambjorn / Jurkiewicz (2005).

[42] Cf. Cremmer / Julia / Scherk (1978).

[43] See section 3.3.1.

Therefore, if gravity should indeed be an emergent, intrinsically classical, macroscopic phenomenon, and not a fundamental interaction, it would not have to be quantized to make it compatible with Quantum Mechanics. Resulting as a classical phenomenon from a quantum substrate, it would already by compatible with Quantum Mechanics. Moreover, it would not only be unnecessary to quantize gravity –it would rather be completely nonsensical to try to quantize gravity. A quantization of gravity would be a quantization of collective, non-fundamental, emergent, macroscopic degrees of freedom. A quantization of General Relativity would be the quantization of an effective theory describing the dynamics of these collective degrees of freedom. It would be as useful as a quantization of the Navier-Stokes equation of hydrodynamics. The resulting 'theory of Quantum Gravity' would be analogous to something like 'Quantum Hydrodynamics': an artificial, formal quantization of a classical theory describing collective, macroscopic degrees of freedom, without any implications for, or any clarifications with regard to, an underlying quantum substrate. It would be simply the wrong degrees of freedom, which are quantized.

So, the option that gravity could be an emergent, intrinsically classical phenomenon would explain very well the problems of all attempts to quantize gravity: conceptual problems as well as those with the reproduction of an adequate classical limit. A quantization of gravity is only (but not necessarily) a reasonable strategy for the construction of a theory of Quantum Gravity if gravity is a fundamental interaction. If it is not a fundamental interaction, the adequate strategy for the development of a theory of Quantum Gravity – then understood primarily as a theory that would dispel the only apparent incompatibility between General Relativity and Quantum Mechanics –consists in the search for the quantum substrate, and for a theory that would explain how the dynamics of the quantum substrate leads to an emergent level with gravitational degrees of freedom.

But, what about spacetime? – If gravity should be an intrinsically classical, residual or induced, emergent phenomenon, without any quantum properties, and if General Relativity gives an adequate description of this intrinsically classical phenomenon, the general relativistic relation between gravity and spacetime, i.e. the geometrization of gravity, should be taken seriously, at least as long as no better reasons make this questionable. General Relativity would have to be seen as a classical, low-energy, long-distance limit to a searched-for theory describing the quantum substrate from which gravity *and* spacetime results. The substrate itself would neither contain gravity, nor would it presuppose spacetime, at least not the continuous, dynamical spacetime of General Relativity[45] into which the gravitational field is encoded as metric field. The spacetime of General Relativity – we would have to expect – would be, like gravity, an emergent phenomenon. It would not be fundamental, but the macroscopic result of the dynamics of a non-spacetime ('pregeometric'[46]) substrate.

---

[44] As a first idea with regard to the emergence of gravity, one could think possibly of an analogy to the emergence of *Van der Waals forces* from electrodynamics.

[45] As mentioned before (section 2), there is already a convincing argument for the existence of discrete microscopic degrees of freedom below the level of a continuous spacetime. It comes from the *Bekenstein-Hawking entropy* of black holes (Cf. Bekenstein (1973, 1974, 1981, 2000, 2001), Wald (1994, 2001), Bousso (2002)) – the paradigmatic 'element of transition'. The *Covariant* (or *Holographic*) *Entropy Bound* (Cf. Bekenstein (1981, 2000, 2001), Bousso (2002), Pesci (2007, 2008)), which can be motivated within the thermodynamics of black holes, can be seen as an indication for an only finite information content of any finite spacetime volume: a finite number of degrees of freedom within a spacetime region. This is in direct contradiction to a continuous spacetime and to the idea of fields defined on this continuous spacetime, fields that imply an infinite number of degrees of freedom for any spacetime region. – This argument for a finite information content of any finite spacetime region can be read as an indication either for a discrete spacetime structure or for a finite pregeometric structure of microconstituents from which spacetime results. The first alternative, that spacetime has a discrete quantum substructure, i.e. that spacetime has quantum properties leading to a finite information content, finds one of its best realizations in the spin networks at the kinematical level of *Loop Quantum Gravity* (see section 3.1.2.). But the, at best, only very limited success of the attempts to quantize gravity and spacetime makes this first alternative less probable. So, the best explanation for the finite information content can be seen in the second

However, if gravity and spacetime should be emergent phenomena, from which structure do they emerge? Of what entities and interactions does the substrate consist?[47] Does matter (and do other quantum fields) also emerge from the substrate? – Meanwhile, there exist a lot of different, more or less (mostly less) convincing scenarios that try to answer these questions; some are conceptually interrelated and some are completely independent. Some of these scenarios take General Relativity as an adequate description of gravity *and* spacetime – as an effective theory for the macroscopic, low-energy regime –, keep to the general relativistic relation between gravity and spacetime, and treat them as emerging together from a pregeometric substrate. Others take General Relativity as a theory with limited validity, even for the classical, macroscopic regime – especially with regard to its geometrization of gravity –, and try to describe the emergence of gravity from a substrate that already presupposes spacetime. Some are pregeometric with regard to space, but not with regard to time, which is presupposed, either as a continuous parameter, or in form of discrete time steps. Most of the scenarios presuppose the validity of Quantum Mechanics on the substrate level, but a few try also to explain the emergence of Quantum Mechanics from a (sometimes deterministic) pre-quantum substrate.[48] – Here is a selection of conceptual ideas that imply that a direct quantization of General Relativity is futile:[49]

### 3.2.1. *Space(time) as an Expression of a Spectrum of States of Pregeometric Quantum Systems*

In the scenario of Kaplunovsky and Weinstein[50] (which does not even mention gravity), space and its dimensionality and topology are dynamical results of the formation of higher-level order parameters within the spectrum of states constituting the low-energy regime of relatively simple pregeometric quantum systems. Fermionic degrees of freedom lead to a flat space; bosonic degrees of freedom lead to a rolled-up space. Besides the geometric order parameters, residual gauge degrees of freedom are typical for the low-energy regime. Ultimately, the distinction between 'geometric' and 'internal' degrees of freedom can be seen in this scenario as a low-energy artifact that has only phenomenological significance. Space is finally nothing more than a fanning-out of a quantum mechanical state spectrum. It is the expression of a quantum system having a low-energy state spectrum that shows a phenomenology that can be interpreted best in a geometrical way. The quantum system, originally pregeometric, has – so to say – a geometric low-energy phase. And possibly not only one: Phase transitions between spaces with different dimensionality are to be expected.

But, the Kaplunovky-Weinstein scenario, based on standard Quantum Mechanics, presupposes an external time parameter, which is at least incompatible with General Relativity. However, first ideas with regard to the question how a temporal dynamics could emerge from a timeless 'dynamics' are arising.[51]

---

alternative; it would then to be read as an indication for a (with regard to its degrees of freedom) finite pregeometric microstructure from which spacetime emerges.

[46] 'Pregeometric' does not necessarily mean 'non-geometric', but 'pre-general-relativistic-spacetime-continuum'.

[47] Certainly it won't be fields, because they presuppose an infinite information content as well as a continuous spacetime on which they are defined.

[48] See also section 4.1.

[49] A more serious pregeometric approach to Quantum Gravity can be found in section 3.4.2.

[50] Cf. Kaplunovsky / Weinstein (1985); see also Dreyer (2004).

[51] Cf. Girelli / Liberati / Sindoni (2008).

### 3.2.2. *Spacetime and Gravity as Emergent Thermodynamic or Statistical Phenomena*

Jacobson[52] has shown that the Einstein field equations can be derived from a generalization of the proportionality between entropy and horizon area for black holes (Bekenstein-Hawking entropy). For that, one needs the thermodynamical relations between heat, temperature and entropy. Temperature has to be interpreted as Unruh temperature of an accelerated observer within a local Rindler horizon. Heat is to be interpreted as energy flow through a causal horizon in the past, leading to a curvature of spacetime, corresponding to a gravitational field. The fundamental dynamics behind the causal horizon, from which the energy flow results, is unobservable in principle, and therefore unknown. Knowledge about this fundamental dynamics is not necessary for the thermodynamical derivation of the Einstein equations. They are generic under equilibrium conditions. Nothing about the fundamental dynamics can be inferred from them.

This seems to be a fundamental problem for many of the emergent gravity / emergent spacetime scenarios. If the Einstein equations are generic (or result at least from entire universality classes of microscopic dynamics that contain completely different dynamical structures), it could be difficult or even impossible to identify reliably the substrate from which gravity and spacetime actually result, at least as long as General Relativity – the macroscopic, low-energy limit – constitutes the complete basis of inference to the substrate.[53] For a reliable identification of the substrate, more information would be necessary, e.g. in form of implications of the matter content of the universe – relevant especially for the identification of the substrate in the context of pregeometric scenarios that couple the emergence of gravity and spacetime intrinsically to the emergence of matter.[54]

### 3.2.3. *Gravity and/or Spacetime as Emergent Hydrodynamic or Condensed Matter Phenomena*

Hydrodynamic and condensed matter models for emergent gravity go back to (and are partially inspired by) Sakharov's *Induced Gravity* scenario[55] of the sixties, which takes gravity as a residual effect of electromagnetism, induced by quantum fluctuations. According to this model, gravity results from Quantum Electrodynamics in the same way as hydrodynamics results from molecular physics. The Einstein-Hilbert action of General Relativity would be an approximate implication of the effective action of a Quantum Field Theory. – More recent approaches, like Hu's model[56], try to reconstruct spacetime as a collective quantum state of many microconstituents forming a macroscopic quantum coherence, comparable to a Bose-Einstein condensate.

The probably most advanced of the condensed matter inspired emergent gravity scenarios is that of Volovik.[57] According to his model, gravity and spacetime could be emergent phenomena resulting from excitation states of a fermionic system with Fermi point (i.e. a topological defect in momentum space). These systems belong to a universality class showing low-energy behavior that reproduces the phenomenology of gravitation, as well as dynamical structures similar to those of the Standard Model of Quantum Field Theory. They contain chiral fermions as low-energy quasi-parti-cles as well as collective bosonic excitation states of the Fermi quantum liquid, which both perceive the condensed matter ground state as an effective frictionless vacuum and its inhomogeneities as metric field, leading thereby to effective gravitational and gauge fields with their corresponding symmetries.

---

[52] Cf. Jacobson (1995, 1999), Eling / Guedens / Jacobson (2006), Jacobson / Parentani (2003). See also Padmanabhan (2002, 2004, 2007).
[53] See also the following sections 3.2.3. and 3.2.4.
[54] See section 3.4.2.
[55] Cf. Sakharov (2000). See also Visser (2002), Barcelo / Liberati / Visser (2005), Weinfurtner (2007).
[56] Cf. Hu (2005). See also Hu / Verdaguer (2003, 2004, 2008), Oriti (2006).
[57] Cf. Volovik (2000, 2001, 2003, 2006, 2007, 2008). See also Finkelstein (1996), Zhang (2002), Tahim et al. (2007), Padmanabhan (2004), Eling (2008).

Again, unfortunately, the exact identification of the substrate – one of the main objectives of a theory of Quantum Gravity – is difficult within Volovik's condensed matter approach to emergent gravity. The best one can achieve is the identification of a universality class from which the known low-energy phenomenology can be reproduced. But such a universality class contains, in general, completely different dynamical systems, which all lead to the same low-energy phenomenology.

In the Fermi-point model, the emergent, effective spacetime is naturally four-dimensional and can have curvature, black holes and event horizons.[58] But the equivalence principle and the general covariance of General Relativity are only approximately valid. Volovik's idea is that this is not necessarily a weakness of the theory. Possibly General Relativity contains theoretical artifacts without counterparts in reality. Its diffeomorphism invariance, representing the general covariance of the theory, could – according to Volovik[59] – be such an artifact, ultimately going beyond the empirically tested phenomenology of gravity.

Actually, it is unclear at the moment, to what extent the hydrodynamic and condensed matter models are in conflict with basic conceptual implications of General Relativity, e.g. what kind of background they need, and if they necessarily need an external time parameter or a quasi-local change rate. Could the background-independence of General Relativity, finally, be just a theoretical artifact, as some of the emergent gravity scenarios suggest? Could, finally, gravity be emergent, but spacetime fundamental? – For an emergent gravity model, a possible background-dependence would at least be less problematic than for an approach starting from a direct quantization of General Relativity (as long as there is no conflict with known phenomenology). In the direct quantization approach, background-dependence would be a conceptual contradiction: a background-dependent quantization of a background-independent theory. For emergent gravity there could still be reasons to take the background-independence of General Relativity as a theoretical artifact. But one really should have very good reasons for this assumption.

### 3.2.4. *Spacetime and Gravity as Phenomenological Results of a Computational Process*

One of the advantages of the idea that spacetime could be an emergent information-theoretical phenomenon is that some of the problematic implications of the hydrodynamic and condensed matter models, e.g. their possible inability to achieve background-independence, can be avoided. The information-theoretical emergent gravity / emergent spacetime scenarios are almost automatically background-independent.[60] – But many alternative scenarios with different substrate constructions exist. Most[61] presuppose quantum principles, but some[62] start from a non-quantum substrate and try not only to elucidate the emergence of gravity and spacetime, but also to reconstruct Quantum Mechanics as an emergent phenomenon.[63]

---

[58] Volovik's model leads – like Hu's – to a natural explanation for a small cosmological constant, as well as for the flatness of the universe.
[59] Cf. Volovik (2007), p.6.
[60] See section 3.4.
[61] Cf. Lloyd (1999, 2005, 2007), Hsu (2007), Livine / Terno (2007), Zizzi (2001, 2004, 2005), Hardy (2007).
[62] Cf. Cahill (2002, 2005), Cahill / Klinger (1996, 1997, 1998, 2005), Requardt (1996, 1996a, 2000).
[63] See section 4.1.

The idea that spacetime emerges from a purely information-theoretical pregeometric substrate goes back to Wheeler's *It from bit* concept[64]. Lloyd[65] modifies this in his *Computational Universe* approach to an *It from qubit*: Spacetime is here to be reconstructed as an emergent result of a completely background-independent quantum computation[66] – a background-independent quantum computer. And because of the background-independence of the substrate, emergent spacetime fulfills – as Lloyd suggests[67] – necessarily the Einstein field equations in their discrete form as Einstein-Regge equations. But, as in almost all emergent gravity / emergent spacetime scenarios, the concrete substrate dynamics, finally, remains obscure. For the *Computational Universe* approach this means: It is unknown, on which concrete computation our universe with its specific spacetime chronogeometry is based.

## 3.3.    Quantum Gravity without the Quantization of General Relativity

It can not be emphasized enough that the most essential requirement for any approach to a theory of Quantum Gravity consists – besides conceptual consistency and coherence – in its ability to reproduce General Relativity (or at least its phenomenology) as a classical limit or a low-energy approximation (up to the exactitude of the already existing empirical data). If no theory that can be constructed by means of a (direct) quantization of General Relativity should be able to fulfill this requirement, two alternative options remain: One could either try to find a quantum theory with the appropriate classical limit by means of the quantization of another classical theory instead of General Relativity. Or one could try to construct or to find such a theory 'directly' – without any quantization of a classical theory at all.

### 3.3.1.  *Quantization of a Classical Theory different from General Relativity as a Road to Quantum Gravity – String Theory*

Without any initial intention to lead to this result, *String Theory*[68] surprisingly turned out to be an instantiation of the first of these two options. Starting from a development in the context of the hadron physics of the sixtieth, where it turned out to be unsuccessful, or at least less successful than Quantum Chromodynamics and the quark model, *String Theory* experienced a sudden reincarnation as a candidate theory for Quantum Gravity, triggered by the observation that the quantization of the classical (special relativistic) dynamics of a one-dimensionally extended oscillating object (the *string*) leads to a perturbative quantum theory containing spin-2 bosons, completely useless in hadron physics. After a shift of the energy level of the oscillating string (the string tension) to the Planck level, and after the elimination of various intratheoretical anomalies, it was possible to reproduce, at least formally, the Einstein field equations as a classical limit.[69] But – forced by the anomaly elimination procedures – that was only possible if one assumed a ten-dimensional spacetime for the string dynamics.

---

[64] Cf. Wheeler (1979, 1983, 1989).

[65] Cf. Lloyd (2005). See also Lloyd (1999, 2007).

[66] Quantum computations are superpositions of computational histories. The transition from these superpositions to a classical macroscopic spacetime consists in their decoherence.

[67] Cf. Lloyd (2005), p.7.

[68] Cf. Polchinski (2000, 2000a), Kaku (1999), Green / Schwarz / Witten (1987). For a survey of the literature, see Marolf (2004).

[69] That it is possible to reproduce the Einstein equations from *String Theory* does not necessarily mean that it reproduces General Relativity in a full-blown sense.

So, *String Theory* does not start from a direct quantization of General Relativity, but instead from the quantization of the classical dynamics of a relativistic string. And spin-2 bosons that can be interpreted as *gravitons* turn out to be quantum states of this string. These graviton states move on a fixed background spacetime, like in the *Covariant Quantization* approach[70]. But *String Theory* seems to evade – obviously with much more success – the problem of the non-renormalizability of the *Covariant Quantization* scheme by means of a nomological unification of all interactions. The oscillation spectrum of the string turned out to contain not only spin-2 states, but – under appropriate conditions that lead to different formulations of *String Theory* – also scalar states, spin-1 gauge bosons and fermionic matter states. Obviously there is not only the exchange boson for gravity, but also states that can possibly be identified – at least formally – with the interaction bosons of the standard model and with its matter particles/fields. And it seems to be the existence of all these string oscillation states that makes the perturbative expansions not only renormalizable, but even finite. Different divergent contributions to the expansion seem to cancel out each other. But all this does only work if one assumes supersymmetry (i.e. the symmetry between bosonic and fermionic states) for the string dynamics. Only if the dynamics of the string fulfills supersymmetry, perturbative *String Theory* seems to be a consistent theory.

However, although they obviously evade the non-renormalizability problem of the *Covariant Quantization* approach, all known formulations of *String Theory* are background-dependent. Instead of simply describing the dynamics of (elementary) gravitons on a fixed spacetime, they describe – simply – the dynamics of one-dimensionally extended strings (with graviton oscillation states) on a fixed spacetime. Without any further reasonable motivation, this is conceptually inadequate for a theory that claims to describe the quantum properties of gravity and spacetime, and to reproduce General Relativity as a classical limit. Without any further idea how a background-independent classical limit could result from a background-dependent quantum theory, the relation between the two theories remains on an exclusively formal level. Without any better reason, a quantum theory leading to General Relativity as a classical limit should be a background-independent theory. The problem was already acknowledged by string theorists; they try to develop a background-independent, non-perturbative formulation of *String Theory* – without success to this day.

And there are a lot of further internal (intratheoretical, conceptual) and external (real physical) problems.[71] One of the internal problems results from the fact that there are (at least) five distinct perturbative string theories – and not only one. By means of relations (dualities) between these five perturbative theories, string theorists try to establish a non-perturbative framework that – at the same time – should possibly solve the problem of the background-dependence. But, as yet, no consistent non-perturbative, analytical framework exists – not even a background-dependent one. A further internal problem consists in the fact that perturbative string theories have necessarily to be supersymmetric, which is in obvious conflict with phenomenology. We do not see supersymmetry in our world, at least not an unbroken one. So, the theory should be able to explain why we do not see supersymmetry, although it is a necessary ingredient of *String Theory*. And there should be numerical predictions with regard to an obviously broken supersymmetry. But *String Theory* has a lot of problems with a broken supersymmetry. And it does not lead to any quantitative predictions at all. Moreover, the problems with regard to the breaking of supersymmetry seem to be coupled to another internal problem: to explain the transition from the necessarily ten-dimensional dynamics of the string (forced by internal consistency requirements) to the four-dimensional phenomenology of our world. Different proposals for dimensional reduction and compactification mechanisms exist. But even if one takes only one compactification scheme into account, this transition is highly ambiguous; it leads to a plethora of four-dimensional low-energy scenarios with different symmetries, oscillation spectra (boson and fermion spectra), etc.: the *string landscape*. Although the string land-scape consists of $10^{500}$ or more four-dimensional scenarios (theories?, models?), it was not possible to identify at least one resembling or reproducing the low-energy phenomenology of our world, or the dynamical structure (and the symmetries) of the standard model respectively. And, unsurpris-ingly, there are simply no numerical predictions at all with regard to the masses of the bosonic and fermionic states of the string – not even for one of the many, many string scenarios.

---

[70] See section 3.1.1.
[71] For a further discussion of *String Theory* and its problems, see Hedrich (2002, 2002a, 2006, 2007, 2007a).

All in all, one has to emphasize that *String Theory* is not supersymmetric because we obviously live in a supersymmetric world. And it does not describe the dynamics of the string in a ten-dimensional spacetime because we obviously live in a ten-dimensional world. It is supersymmetric and ten-dimensional because it is not possible to find a formulation of the theory without these features. Both features are exclusively internally motivated, by means of mathematical consistency require-ments resulting from the decision to quantize the dynamics of a relativistic string. They do not have any external, truly physical motivation. Even after more than three decades of development, there does not exist the slightest idea with regard to a fundamental *physical* principle on which *String Theory* should be based or from which it could be motivated or developed. The only motivation for *String Theory* consists still in the post-hoc discovery that a quantization of a relativistic string leads to spin-2 oscillation states that can be identified with gravitons. But it can be shown that (almost) every theory describing the dynamics of spin-2 states reproduces formally the Einstein equations. So, if the theory remains background-dependent (without any motivation for that), if it leads addi-tionally to the contingency problem related with the string landscape, and if it has no true physical motivation going beyond its post-hoc discovery of graviton states, this latter discovery is certainly not enough to take the theory seriously as an adequate description of the quantum substrate of grav-ity and spacetime.

### 3.3.2.  *Quantum Gravity without Quantization of any Classical Theory*

Taken into account the problems of *String Theory*[72] (still the most popular approach to Quantum Gravity, not at least with regard to the number of investigators involved) as well as those of *Loop Quantum Gravity*[73] (the most promising variant of the direct, canonical quantization approach), it seems reasonable to take the quest for alternatives seriously. The direct quantization of a classical theory – whether General Relativity or some other classical dynamics – is not necessarily the only way that could possibly lead to a theory of Quantum Gravity. The main objective of such a theory consists in the identification of the substrate leading to macroscopic gravity (and spacetime), irre-spective of the question by means of which method this goal can be achieved. Meanwhile, the spectrum of alternatives to the direct quantization approaches and of their specific backgrounds and motivations is quite varied.[74] And it is especially the field of pre-spacetime ('pregeometric') ap-proaches to Quantum Gravity that could possibly offer perspectives to avoid the problems resulting for the direct quantization approaches:

---

[72] In summary: After more than three decades of development, *String Theory* remains a perturbative construct without any physically motivated fundamental principle. Like Quantum Field Theories, it works with a fixed background spacetime, inconsistent with General Relativity. In addition, since the discovery of the string landscape, the perspectives with regard to the predictive power of the theory are almost hopeless.

[73] In summary: *Loop Quantum Gravity* has severe conceptual problems related to the quantum Hamiltonian constraint. The perspectives for a reproduction of the known phenomenology of gravity, or of General Relativity as a classical, macroscopic, low-energy limit, are completely unclear.

[74] See also section 3.2.

### 3.4. Pregeometric Theories

#### 3.4.1. *Motivations and Intuitions*

All pregeometric theories have in common that they describe some kind of dynamics of discrete structures that do not presuppose (continuous) spacetime (i.e. a spacetime manifold). The basic assumption of pregeometric theories is that spacetime and gravity are emergent phenomena, resulting from the dynamics of a substrate without any gravitational or continuous-spacetime degrees of freedom. The indispensable requirements for theories of Quantum Gravity hold without any modification for pregeometric theories; in particular, they have necessarily to reproduce General Relativity (or at least its phenomenology) as an approximation or as a macroscopic continuum limit.

The most important motivation for these approaches – despite their differences – can be seen in the numerous clues pointing to a discrete substrate of spacetime, to be found within the (extended) context of our well-established theories (especially the *Bekenstein-Hawking entropy* as a result of the interplay of arguments from General Relativity, Quantum Field Theory, and Thermodynamics) as well as coming from the field beyond these established theories (discrete spin nets in *Loop Quantum Gravity*; indications of a minimum length in *String Theory*; the (formal) reproduction of the *Bekenstein-Hawking entropy* in both approaches).

Meanwhile, there exists a considerable variety of different pregeometric approaches to Quantum Gravity: *Causal Sets*[75], computational scenarios[76], *Pregeometric Quantum Causal Histories*[77], etc.[78] They differ not at least with regard to their specific construction of the substrate dynamics from which spacetime and gravity is supposed to emerge. Most of the approaches presuppose Quantum Mechanics[79] and start with elementary (quantum) events without any spacetime embedding as basic elements on the substrate level. These elementary events are the nodes (vertices) of a network of equivalent (quantum) relations that can be modeled by a graph. They differ especially with regard to the interpretation of these relations between the elementary events: some take them as a representation of basic causal relations between elementary events (e.g. *Causal Sets*), others as dynamical quantum information channels (e.g. the computational approaches); the *Quantum Causal Histories* approach declares both options as identical.[80]

But, independently of the problem of the identification (or construction) of the substrate, the question remains: How can spacetime emerge from something so completely different from spacetime as quantum information, information flow, or basic causal relations? How can the chronogeometry of spacetime emerge from something completely pregeometric? This is probably one of the most fundamental questions to be posed with regard to the pregeometric scenarios that try to describe the emergence of spacetime. The question results from the obvious conflict of these scenarios with our intuitions about spacetime. – A possible reconciliation with our intuitions comes from the *Holographic Screens* idea: [81] Take an acyclic network (a graph) of directed relations ('lines') between elementary quantum systems ('vertices') without any (continuous, metrical) spacetime background. The directed relations are instantiated by flows of quantum information between the elementary

---

quantum systems (and can be interpreted as causal relations). Dynamical changes occur locally in discrete steps. There are no continuous spacetime degrees of freedom on the fundamental level. – Then define screens that separate adjacent parts of the relational network, cutting through some of the lines of the network. For each screen a specific quantum information flow capacity can be found. The crucial idea of the *Holographic Screens* concept starts from an inversion of the central implications of the Bekenstein-Hawking entropy:[82] According to Bekenstein, the entropy of a black hole is proportional to the area of its event horizon. And, according to the Holographic (or Covariant) Entropy Bound,[83] this Bekenstein-Hawking entropy defines the maximum information content of the corresponding volume. The maximum information (corresponding to the number of independent degrees of freedom) contained within a spacetime volume is finite and proportional to the area of the surface of the spacetime volume. The inversion of this Holographic Bound – the core of the *Holographic Screens* concept – consists now in the idea that the amount of quantum information that can flow through a screen (the quantum information flow capacity of the screen) defines the *area* of the screen. And then, after having defined area as information flow capacity, a spacetime geometry can be established by means of a (secondary) network of 'holographic screens', to be defined on the (primary) network of elementary quantum systems and their causal relations.[84] So, the *Holographic Screens* concept exemplifies, how Wheeler's *It from bit*[85] – modified to an *It from qubit* – could work in principle.

### 3.4.2. *Quantum Causal Histories*

The intuitions developed with the *Holographic Screens* idea can be seen as an ideal starting point for an approach that goes by the name *Pregeometric Quantum Causal Histories*[86]. This theoretical approach is at the moment probably the most general and at the same time the most clear-cut, paradigmatic case[87] of an attempt to construct a theory of Quantum Gravity that can explain how gravity as well as spacetime – here both have no quantum properties, because they are intrinsically classical phenomena – could emerge from a very simple pregeometric substrate based exclusively on quantum information and its flow. Although the approach is at the moment completely speculative,[88] it shows especially how to avoid the most prominent problems of the direct quantization approaches, as well as those of most other emergent gravity / emergent spacetime scenarios.[89] – Its basic assumptions are:

---

[82] Cf. Bekenstein (1973, 1974, 1981, 2000, 2001), Wald (1994, 2001), Bousso (2002). See also section 2.

[83] Cf. Bekenstein (1981, 2000, 2001), Bousso (2002), Pesci (2006, 2007). See also section 2.

[84] The flow of quantum information through holographic screens defines not at least the causal structure of spacetime. Compare also Jacobsons thermodynamic approach to an emergent spacetime. See section 3.2.2.

[85] See section 3.2.4.

[86] Cf. Markopoulou (2000, 2000a, 2000b, 2004, 2006, 2007), Dreyer (2004, 2006, 2007) (Dreyer calls his approach *Internal Gravity*), Kribs / Markopoulou (2005), Konopka / Markopoulou / Smolin (2006) (*Quantum Graphity*), Konopka / Markopoulou / Severini (2008), Hawkins / Markopoulou / Sahlmann (2003).

[87] *Quantum Causal Histories* can not only be seen as the paradigmatic case of a pregeometric theory of Quantum Gravity, but also as a synthesis or a point of convergence of many different approaches to a pregeometric quantum substrate. They are, on the one hand, a conceptual extension of Sorkin's *Causal Set* approach (Cf. Bombelli / Lee / Meyer / Sorkin (1987), Sorkin (2003), Rideout / Sorkin (2000, 2001), Rideout (2002), Henson (2006), Surya (2007)), enriched by the *Holographic Screens* idea (Cf. Markopoulou / Smolin (1999)) and elements from Lloyd's *Computational Universe* scenario (Cf. Lloyd (1999, 2005, 2007)), which itself owes a lot to Wheeler's *It from bit* (Cf. Wheeler (1989)). On the other hand, *Quantum Causal Histories* can also be seen as a generalization of causal spin networks and of the *Spin Foam* approach (Cf. Oriti (2001, 2003), Livine / Oriti (2003), Perez (2003, 2006), Baez (1998, 2000), Markopoulou / Smolin (1997)), enriched by elements from *Algebraic Quantum Field Theory*.

[88] To be exact, there is, as yet, no theory of Quantum Gravity that is not completely speculative.

[89] See section 3.4.2.

- – Causal order is more fundamental than properties of spacetime, like metric or topology.
- – Causal relations are to be found on the substrate level in form of elementary causal network structures.
- – There is no continuous spacetime on the substrate level. The fundamental level does not even contain any spacetime degrees of freedom at all.
- – Only a finite amount of information can be ascribed to a finite part of the substrate network of causal relations.[90]
- – Quantum Mechanics is valid on the fundamental level.

*Quantum Causal Histories* are relational networks of quantum systems with only locally defined dynamical transitions. The basic structure is a discrete, directed, locally finite, acyclic graph. To every vertex (i.e. elementary event) of the graph, a finite-dimensional Hilbert space (and a matrix algebra of operators working on this Hilbert space) is assigned.[91] So, every vertex is a quantum system. Every (directed) line of the graph stands for a causal relation: a connection between two elementary events; formally it corresponds to a quantum channel, describing the quantum evolution from one Hilbert space to another. So, the graph structure becomes a network of flows of quantum information between elementary quantum events. *Quantum Causal Histories* are information processing quantum systems; they are quantum computers.

Because there are no spacetime degrees of freedom on the fundamental level of description, *Quantum Causal Histories* are necessarily background-independent, and therefore not in direct conceptual conflict with General Relativity. However, if this approach intends to be successful as a theory of Quantum Gravity, it has to explain *geometrogenesis*; it has to explain how spacetime emerges from the pregeometric quantum substrate. This would be the first step on the way to a reproduction of the empirically well-tested phenomenological implications of General Relativity – the most basic and indispensable requirement for any theory of Quantum Gravity. But it is not enough: General Relativity itself has to be reproduced as an effective theory for the macro-level. Therefore, a second step should consist in the explicit reproduction of the Einstein field equations as a classical, macroscopic approximation.

The basic idea of the *Quantum Causal Histories* approach with regard to the first step – geometrogenesis – is the following: Macroscopic spacetime and classical gravity do not result from a coarsegraining of quantum-geometric degrees of freedom – those do not exist according to the *Quantum Causal Histories* approach –, but from the dynamics of propagating coherent excitation states of the substrate.[92] Then, macroscopic spacetime is necessarily dynamical because it results from a background-independent pregeometric dynamics.[93] The effective degrees of freedom on the macro-level are necessarily decoupled from the dynamics of the substrate degrees of freedom. If they were not, there would not be any spacetime or gravity on the macro-level, because there is none on the substrate level. In the same way, causality on the macro-level, finding its expression in the macro-level interactions, is decoupled from causality on the substrate-level. Spacetime-locality on the macrolevel, if it emerges from the dynamics of coherent excitation states, has nothing to do with locality on the substrate graph structure level.[94]

---

[90] This assumption is motivated explicitly by the Bekenstein-Hawking entropy (and the Holographic Entropy Bound) which leads to finite information limits for finite regions, and which can be reproduced under certain conditions even by *Loop Quantum Gravity* (cf. Meissner (2004)) and by *String Theory* (cf. Das / Mathur (2001), Lemos (2005), Peet (1998, 2001), Maldacena (1996)).

[91] This is one of the most important extensions in comparison with the *Causal Set* approach.

[92] Cf. Kribs / Markopoulou (2005).

[93] However, not every pregeometric substrate has necessarily a geometric phase.

[94] Cf. Markopoulou (2006) pp. 24f.

But what are these coherent, propagating excitation states, resulting from the substrate dynamics and leading to spacetime and gravity? And *how* do they give rise to spacetime and gravity? – The answer given by the *Quantum Causal Histories* approach consists in a coupling of geometrogenesis to the emergence of matter. The idea is that the coherent excitation states resulting from (and at the same time dynamically decoupled from) the substrate dynamics are matter degrees of freedom. And they give rise to spacetime, because they behave as if they were living in a spacetime. The space-time of the *Quantum Causal Histories* approach is nothing more than an implication of the behavior of (emergent) matter. Spacetime is a completely relational construct, an expression of the phenome-nology of matter dynamics. – And the matter degrees of freedom give at the same time rise to gravity, because the spacetime they bring forth by means of their behavior is a curved spacetime.[95] Gravity is nothing more than an expression of this curved spacetime.[96] The still unproved central hypothesis of the *Quantum Causal Histories* approach is that the Einstein field equations are neces-sarily an implication of the dynamics of these coherent excitation states, and that they can finally be derived from the substrate dynamics.[97]

But, what kind of matter does emerge from the substrate of the *Quantum Causal Histories* ap-proach? And what is it that stabilizes the coherent excitation states corresponding to matter? – The answer to the last question is: topology. The idea is that the coherent excitation states can be identi-fied with stable topological knot structures: braids with crossings and twists.[98] These topological structures seem to be conserved by the substrate dynamics because of topological symmetries, i.e. because of the corresponding topological conservation principles. And, interestingly, the basic properties of these stable topological structures can be identified with the well-known basic proper-ties of elementary particles.[99] All particles of the Standard Model can be identified with specific topological structures.[100] – However, what is still missing, is a dynamical explanation that elucidates more extensively the identification of the basic properties of the stable topological structures with the basic physical properties of elementary particles. It should, finally, be possible to derive energy conservation principles from the dynamics of the stable topological structures (which should be translation-invariant); and this should, not at least, lead to an explanation for particle masses.

## 4.    Beyond Quantum Gravity

### 4.1.    A Fundamental Theory without the Quantum ?

The question remains if the already mentioned approaches to a theory of Quantum Gravity, finally, are sufficient (or radical enough) to get over the (possibly only apparent) conceptual incompatibility between General Relativity and Quantum Mechanics / Quantum Field Theory, and if at least one of these approaches has the potential to attain at the same time an empirically adequate description of nature, consistent also with future empirical data that go beyond those on which the established theories are based. The more orthodox, mainstream approaches to Quantum Gravity, like *String Theory* and *Loop Quantum Gravity*, seem to lead to severe conceptual problems and are unable, at least at the moment, to reproduce the phenomenology of our established, but apparently mutually incompatible theories. And the less orthodox approaches – especially the emergent gravity / emergent spacetime scenarios, like the *Quantum Causal Histories* approach – are at the moment only more or less developed conceptual ideas, far from a full theoretical framework. Almost all of these approaches, orthodox or less orthodox, presuppose Quantum Mechanics. They either suppose that there are quantum properties of gravity and spacetime, or they start with a quantum substrate from which gravity and spacetime result as emergent, intrinsically classical phenomena.

---

[95] There are already concrete indications in *Quantum Causal Histories* for a curved spacetime with Lorentz signature.

[96] And gravity has, as part of macro-causality, a finite propagation speed, because the coherent excitation states of the substrate, the matter degrees of freedom, have a finite propagation speed.

[97] Cf. Markopoulou (2007) p. 19.

[98] Cf. Bilson-Thompson / Markopoulou / Smolin (2006), Bilson-Thompson (2005).

[99] E.g., the twist of a braid structure can be interpreted as electromagnetic charge. There are also topological counter-parts to charge conjugation, to quark colors, to parity, etc. Cf. Bilson-Thompson / Markopoulou / Smolin (2006), p. 6.

[100] Cf. Bilson-Thompson / Markopoulou / Smolin (2006), Bilson-Thompson (2005), Bilson-Thompson / Hackett / Kauffman / Smolin (2008). Naturally, the spectrum of topological structures does not contain any counterpart to the graviton. According to the *Quantum Causal Histories* approach there are no gravitons. Gravity is an intrinsically classi-cal, emergent phenomenon; it does not have any quantum properties or quantum constituents.

Some people think that such attempts at a construction of a theory of Quantum Gravity are not radical enough, that not only gravity and spacetime, but also the quantum could be an emergent phenomenon.[101] According to those people, the still unknown fundamental theory could quite perfectly be a non-quantum theory, describing a substrate from which gravity, spacetime and the quantum emerge. The probably best-known of these emergent quantum approaches goes back to 't Hooft.[102] He proposes a deterministic, pregeometric, non-quantum substrate, which should possibly be modeled by something like cellular automata.[103] None of these proposals has achieved a concrete theoretical framework so far.

## 4.2.    No Fundamental *Theory* – The Possibility of an Anomological Substrate

There are also approaches that do not call into question especially the universal validity of Quantum Mechanics, but instead that of a fundamental nomologicity of nature in general. The idea is that the laws of nature themselves – and all of them – are emergent, resulting from a lawless substrate, possibly by means of something like a statistical coarse-graining. (Or they are even a consequence of our scientific methodology and its search for regularities and nomological structures.) Then, the laws of nature would be only approximately valid, 'macroscopic', low-energy phenomena. Nature would ultimately, on its most fundamental level, be anomological and chaotic. Best known is Wheeler's idea of a *Law-without-law* physics.[104] In the context of Nielsen's *Random Dynamics*[105] – a concretization of Wheeler's idea – it was even possible to derive some of the physically most important symmetries and regularities as approximately valid lawful structures from a lawless, chaotic substrate.

## 4.3.    No *Fundamental* Theory – Patchwork Physics

Should all attempts to get rid of the conceptual incompatibility between our established fundamental theories – General Relativity and Quantum Mechanics / Quantum Field Theories – remain without success on the long run, the last option would consists in the view that a unified, conceptually coherent physical description of nature can possibly not be achieved. Maybe physical theories can only be seen as theoretical instruments with a limited explicatory scope. Maybe they do not lead to a coherent, unified description of nature. Then all attempts to reach at a fundamental physical description of nature, to reach at a unified theory describing an ultimate substrate dynamics, would probably be conceptually inadequate extrapolations of our nomological ambitions. The assumption of a fundamental unity of nature would be simply wrong, at least as far as it concerns its reflection within our theoretical and methodological apparatus of physics. Maybe we are living in a *Dappled World*[106], consisting of disparate realms of phenomena, each of which makes necessary a different, more or less completely autonomous scientific approach. Maybe there are even sectors of reality that completely block out all scientific endeavors. – Under these conditions, a (successful, empirically adequate) theory of Quantum Gravity would not be achievable. But, before being satisfied with this option, one should have taken into account and tried out all alternatives: known or still unknown, orthodox or as radical as they might be. And under which conditions one could say that this has already been done!

---

[101] Cf. Requardt (1996, 1996a, 1996b, 2000), Cahill (2002, 2005), Cahill / Klinger (1996, 1997, 1998, 2005). See also section 3.2.4.

[102] Cf. 't Hooft (1999, 2000, 2001, 2001a, 2007). See also Suarez (2007).

[103] The approach of Requardt (1996, 1996a, 1996b, 2000) has certain similarities with regard to this.

[104] Cf. Wheeler (1979, 1983).

[105] Cf. Nielsen (1983), Frogatt / Nielsen (1991), Nielsen / Rugh (1994), Nielsen / Rugh / Surlykke (1994), Bennett / Brene / Nielsen (1987).

---

[106] Cf. Cartwright (1999). See also Cartwright (1983, 1989, 1994) and Morrison (2000).